\documentclass[11pt]{article}
\usepackage{graphicx}

\begin{document}

\def\xslash#1{{\rlap{$#1$}/}}
\def \p {\partial}
\def \dd {\psi_{u\bar dg}}
\def \ddp {\psi_{u\bar dgg}}
\def \pq {\psi_{u\bar d\bar uu}}
\def \jpsi {J/\psi}
\def \psip {\psi^\prime}
\def \to {\rightarrow}
\def\bfsig{\mbox{\boldmath$\sigma$}}
\def\DT{\mbox{\boldmath$\Delta_T $}}
\def\xit{\mbox{\boldmath$\xi_\perp $}}
\def \jpsi {J/\psi}
\def\bfej{\mbox{\boldmath$\epsilon$}}
\def\ej{\mbox{$\epsilon$}}
\def \t {\tilde}
\def\epn {\varepsilon}
\def \up {\uparrow}
\def \dn {\downarrow}
\def \da {\dagger}
\def \pn3 {\phi_{u\bar d g}}

\def \p4n {\phi_{u\bar d gg}}

\def \bx {\bar x}
\def \by {\bar y}
\begin{center}
{\Large\bf Spin-Flip Interactions and the Puzzle of $\psi's$ Polarization at Tevatron}
\vskip 10mm
Kui-Yong Liu, Jian-Ping Ma and Xing-Gang Wu     \\
{\small {\it Institute of Theoretical Physics, Academia Sinica,
Beijing 100080, China}}  \\
\end{center}

\vskip 1cm
\begin{abstract}
Nonrelativistic QCD provides a systematic approach for inclusive
decays and productions of a quarkonium. By taking color-octet
components into account, the approach can explain the
$\psi'$-anomaly at Tevatron, where the measured production rate at
large transverse momentum $p_\perp$ is in order of magnitude larger
than the predicted with color-singlet components only.
With the approach one can predict that the produced $J/\psi$ and
$\psi'$ at large $p_\perp$ will be transversely polarized.
But the prediction fails in confronting with experimental measurements
and this generates a puzzle. We examine the
role of spin-flip  interactions in the spin density matrix of the
transition of a color-octet charm quark pair into $J/\psi$ and $\psi'$. These
interactions will introduce new nonperturbative parameters
in the spin density matrix.
Our result shows that the impact of the interactions is always
to dilute the polarization and can
be very significant. Taking the impact into account,
predictions for the polarization are more
close to the measured than the previous predicted.
The same can also be expected for the polarization of $J/\psi$.

\vskip 5mm
\noindent
% PACS numbers:
\end{abstract}
\vskip 1cm
\par\eject
Quarkonia with their rich dynamics at different energy scales
provide a special place to understand QCD. With its heavy mass $m$
a heavy quark $Q$ will move with a small velocity $v$ inside
a quarkonium. This results in a hierarchy of energy scales
$m>> mv >> mv^2$ and the dynamics at different energy scales
is different. A systematical separation of effects from
the dynamics at different energy scales will provide
a systematic understanding of quarkonium physics.
Indeed, nonrelativistic QCD
(NRQCD) provides such a systematic approach for inclusive decay and production
\cite{NRQCD}. In this approach a factorization
can be made by expanding the small $v$ to do the separation, a power counting in $v$
based on the hierarchy is obtained to determine  which nonperturbative effects,
represented by NRQCD matrix elements, are relevant\cite{NRQCD,pv}. A
comprehensive review of quarkonium physics can be found in \cite{Bra}.
In this approach, effects of higher-Fock components of a quarkonium state  can be
taken into account systematically. Although the probability to
find such higher-Fock components is suppressed,
but the effects of these higher-Fock states can be very significant.
This has been shown in the explanation of the $\psi'$-anomaly at Tevatron, where
the inclusive $\psi'$-production rates
with large transverse momentum $p_\perp$ are in order of magnitude larger than
the predicted if one only takes
the main Fock state into account.
By taking color-octet $Q\bar Q$ components into account,
the Tevatron data\cite{CrosEx} can be explained\cite{BrFl}. This is regarded as a great
triumph of the NRQCD approach.
\par
Despite many successes of the approach, some problems still remain unsolved.
Among them a crucial one is that
the approach fails to predict the polarization of $J/\psi$ and $\psi'$ at large $p_\perp$
measured at Tevatron. At large $p_\perp$, the production of $J/\psi$ and $\psi'$
is dominant by the gluon fragmentation in which a gluon fragments into
a color-octet $c\bar c$ pair in $^3S_1$ state, the color-octet $c\bar c$ pair
is then transmitted into $J/\psi$ and $\psi'$.
This color-octet mechanism successfully explains the unpolarized
production rate at large $p_\perp$ at Tevatron. If the spin-symmetry for charm quarks holds,
the $J/\psi$ and $\psi'$ produced through the gluon fragmentation is always
transversely polarized\cite{ChoWise}. Hence, this mechanism also predicts
that $J/\psi$ and $\psi'$ are transversely polarized at large $p_\perp$.
There are some effects which can dilute the polarization,
like feeddown from higher quarkonium states, higher order corrections
in $\alpha_s$ of the gluon fragmentation\cite{BeRo}. Including these effects
there is still a large fraction of $J/\psi$ or $\psi'$
with transverse polarization\cite{Polth1,AKL,BKL}. But this prediction
is in conflict with the measured at Tevatron\cite{Polex}.
This is puzzling because the same mechanism explains the production of unpolarized
$J/\psi$ and $\psi'$ at large $p_\perp$ but fails to explain the polarization of the produced
$J/\psi$ and $\psi'$. This is the well-known puzzle.
\par
Giving the fact that the color-octet mechanism can explain the unpolarized
production at large $p_\perp$, for solving the puzzle one needs to carefully examine the transition
of a color-octet $^3S_1$ $c\bar c$ pair into $\psi$ which stands for $J/\psi$ or $\psi'$, especially
to examine how the polarization of the pair is transmitted.
This is the purpose of the letter. At the leading order of $v$
a color-octet $^3S_1$ $c\bar c$ pair
will be transmitted into a $\psi$ though the electrical dipole($E_1$) interaction.
This interaction conserves the spin of heavy quarks, hence the spin of the $c\bar c$
pair is totally transmitted into the spin of $\psi$. It is also known
that the magnetic dipole ($M_1$) interaction violates
the spin symmetry, this will make a difference between the spin of the $c\bar c$ pair
and that of $\psi$. With the standard power counting the difference is at order of $v^2$,
as will be shown later.
It is known that the suppression by $v^2$
for charmoina is rather weak. Hence the difference can be significant.
The standard power counting is based on the hierarchy
$m>> mv >> mv^2 \sim \Lambda_{QCD}$.
For charmonia, $mv$ is already below $1$GeV, and it is
at the same order of $\Lambda_{QCD}$. This stimulates to modify
the power counting for charmonia\cite{GS,MB,FRL} under the hierarchy
$m>> mv \sim \Lambda_{QCD}$. In the modified
power counting dynamical gluons are only with one typical momentum
whose components are all at order of $\Lambda_{QCD}\sim mv$.
With this power counting, the $E_1$- and $M_1$ interaction are at
the same order $\Lambda^2_{QCD}$ or $v^2$.
Since the effect of spin-flip interactions is weakly suppressed
or is not suppressed, the spin of a color-octet $^3S_1$ $c\bar c$ pair
will be not transmitted into the spin of $\psi$ completely.
We will examine
the role of the spin-flip interactions with these two power-counting methods
and study their impact on the polarization of $\psi'$ produced
at Tevatron.
It turns out that the impact is significant and always
to dilute the polarization of $\psi'$. Hence the puzzle
can be answered, at least at certain level.
\par
We consider an inclusive $\psi$ production through a color octet
$c\bar c$ quark pair in $^3S_1$ state. Assuming the produced $\psi$
is measured with the momentum $p$,
we define the rest frame
of $\psi$ by a Lorentz boost from its moving frame. The produced
$\psi$ is polarized with the polarization vector $\bfej^*$ in
the rest frame. After decomposing Dirac- and color indices in NRQCD factorization,
the contribution to the differential cross section
can be generally written as:
\begin{equation}
 d\sigma [ ^3 S_1^{(8)} ] =  H_{ij}  T_{ij}(\bfej, \bfej^*,\hat{\bf p}),
\end{equation}
where $H_{ij}$ is the $3\times 3$ spin density matrix for producing a $^3 S_1$  $c\bar c$ pair in color-octet.
This matrix can be calculated with perturbative QCD and contains all kinematical information.
For production at a hadron collider, it can be written as a convolution
with parton distributions and perturbative functions.
The matrix $T_{ij}$ is
the spin density matrix for the transition of the $^3 S_1$  $c\bar c$ pair in color-octet
into a polarized $\psi$ and contains all nonperturbative information of the
transition. It is defined as:
\begin{equation}
 T_{ij}(\bfej, \bfej^*,\hat{\bf p}) = \sum_X \langle 0 \vert \chi^\dagger T^b \sigma^i \psi
 L^\dagger_{bc}  \vert \psi(\bfej ) +X \rangle
    \langle \psi(\bfej^* ) +X \vert L_{ca}\psi^\dagger T^a \sigma^j \chi \vert 0 \rangle.
\end{equation}
where the matrix element is defined in the rest frame and all NRQCD
fields are at the origin of the space-time. The field
$\psi(\chi^\dagger)$ annihilates a $c(\bar c)$-quark respectively.
$L$ is a gauge link defined as $P \exp\left( -ig_s \int_0^\infty
d\lambda n\cdot G ( \lambda n )\right )$ with the gauge field in the
adjoint representation of $SU(3)$. $n$ is determined by the moving
direction of $\psi$,
it is $n^\mu = (1, -{\hat{\bf p}})/{\sqrt{2}}$ with ${\hat{\bf p}}={\bf p}/\vert{\bf p}\vert$.
This gauge link should be added, not only
because of the gauge invariance of $T_{ij}$, but also because the
completeness of NRQCD factorization at two-loop level, as shown by
recent works\cite{NQS}. This gauge link has no effect on NRQCD
factorization at one-loop level\cite{NQS, MaSi}.
\par
The matrix $T_{ij}$ can be decomposed as:
\begin{eqnarray}
T_{ij}(\bfej,\bfej^*, \hat{\bf p}) &=& \delta_{ij} \left ( \bfej\cdot \bfej^* a_0
                        + \bfej\cdot\hat{\bf p}\cdot \bfej^*\cdot\hat{\bf p} a_1 \right )
                        +(\ej_i \ej_j^* + \ej_j \ej_i^*) c_0
\nonumber\\
   && +\left [ \left (\ej_i \hat p_j +\ej_j \hat p_i\right )\bfej^*\cdot\hat{\bf p} +(\bfej\leftrightarrow\bfej^*)\right ]
   c_1
   + \hat p_i \hat p_j \bfej\cdot \bfej^*  c_2 +\cdots,
\end{eqnarray}
where $\cdots$ denote the part which is anti-symmetric in $\bfej$
and $\bfej^*$ or in $i$ and $j$. This anti-symmetric part is
irrelevant here. In NRQCD the leading order interaction
has the heavy quark spin symmetry. With the symmetry one has
$a_{0,1}=c_{1,2}=0$, i.e., only $c_0$ is not zero.
This implies that $\psi$ will have the same spin as the color-octet
$c\bar c$ pair does. However, the symmetry holds only approximately.
Hence, in general those coefficients beside $c_0$ are not zero.
With Eq.(3) we have the
total cross-section $d\sigma_{tot}$ and the cross-section
$d\sigma_L$ with longitudinal polarization:
\begin{eqnarray}
  d\sigma [ ^3 S_1^{(8)} ]_{tot} &=& H_{ij} \delta_{ij} \left [ 2 c_0 + 3 a_0 +a_1 \right ]
                  + {\hat p}_i {\hat p}_j H_{ij} \left [ 4 c_1 +3c_2 \right ],
\nonumber\\
  d\sigma [ ^3 S_1^{(8)} ]_{L}  &=&  H_{ij} \delta_{ij} \left [ a_0 + a_1  \right ]
                  + {\hat p}_i {\hat p}_j H_{ij} \left [ 2 c_0 + 4c_1 + c_2 \right ].
\end{eqnarray}
For Tevatron, one can find that the quantity  ${\hat p}_i {\hat p}_j H_{ij}$
is suppressed by $1/p_\perp^2$ relatively to the quantity $H_{ij} \delta_{ij}$ when
$p_\perp$ is large. Therefore, for large $p_\perp$ the first term in the above two cross
sections is dominant. If  $a_{0,1}$ are zero,
$d\sigma_{L}$ is suppressed by $1/p_\perp^2$ in comparison
with $d\sigma_{tot}$ and leads to the prediction that the produced $\psi$ will be transversely
polarized at large $p_\perp$. But, this is not observed in experiment\cite{Polex}.
It should be noted that the first term can also be factorized with the gluon fragmentation
function into $\psi$ and the function can be studied with NRQCD factorization\cite{NQS,gfj}.
\par
From the above discussion, it is clear that $d\sigma_{L}$ will not be suppressed by
$1/p_\perp^2$, if one takes spin-flip interaction into account, i.e., if
those coefficients beside $c_0$ are not zero.
Unfortunately, these coefficients are unknown yet.
The corresponding coefficients of the matrix element relevant to inclusive $\psi$-decays
has been studied with lattice QCD\cite{BLS}. It has been found that the effect of spin-flip
interactions is small. It should be noted that
the matrix element for decays is different than $T_{ij}$ for productions
and the renormalization effect is not
taken into account in \cite{BLS}. This effect can be difficult to study
because one has operators which are power-divergent(see e.g., in \cite{MNS}).
Although one does not know $a_{0,1}$ and $c_{1,2}$, one can determine
their relative importance by a power counting in $v$, where
one replaces the state $\psi$ in $T_{ij}$ with a free $c\bar c$ pair in a color-singlet $^3 S_1$ state
and calculates $T_{ij}$  with perturbative theory to extract those coefficients.
In perturbative theory the intermediate state $X$ in $T_{ij}$ should consist of two gluons at least.
Taking the leading order interaction of NRQCD,
one easily finds that $c_0$ is at order of $v^4$  and other coefficients are zero as expected.
To have nonzero $a_0$, one of the two gluons must be emitted or absorbed by a heavy quark
with spin-flip interaction. Taking the $M_1$ interaction into account, one finds that
$a_0$ is power divergent when the energy of the gluons becomes large.
To have nonzero $a_1$, $c_1$ and $c_2$ an extra gluon exchanged between a heavy quark and the gauge link in $T_{ij}$
is required.
They are also power divergent. It should be noted
that in NRQCD the dynamical freedoms
of gluons with the energy larger than the heavy quark mass are integrated out. Hence
those gluons in the intermediate state can only have energies  below the heavy quark mass.
With a cut-off scale $\Lambda_s$ for the energy of the state
$X$ we have then:
\begin{equation}
 a_0 \sim \alpha_s^2 (\Lambda_s)\Lambda_s^4, \ \ \ \ \ \ \  a_1 \sim c_1 \sim c_2 \sim
 \alpha_s^3 (\Lambda_s)\Lambda_s^4.
\end{equation}
The scale $\Lambda_s$ should be understood as a characteristic scale for energies
of those intermediate gluons.
\par
If we assume that gluons with
the energy at order of $mv$ are dominant in the transition, one should take
$\Lambda_s =mv$. In this case the strong coupling $\alpha_s$ should be taken
as at order of $v$\cite{NRQCD,pv}.  With
$\Lambda_s =mv$
we have
\begin{equation}
 a_0 \sim  v^6 \ \ \ \ \ \ \  a_1 \sim c_1 \sim c_2 \sim
  v^7.
\end{equation}
From the above $a_0/c_0$ is suppressed by $v^2$ and the ratio of other coefficients is
suppressed by $v^3$.  For charmonia corrections at the next-to-leading order of $v$ are
generally not small,
they can be numerically of the same size of the leading contributions
in different processes\cite{ReCo}. Hence those coefficients can have
a significant impact on theoretical predictions.
\par
The power counting given above is the standard one.
As discussed before, a modification of the power counting can be needed because
$mv$ is close to $\Lambda_{QCD}$ for charmonia\cite{GS,MB,FRL}. With the modification
$\alpha_s(\Lambda_s)$ with $\Lambda_s =mv\sim \Lambda_{QCD}$ should be taken as $1$. Then we have:
\begin{equation}
 \frac{a_0 }{c_0}\sim  \frac{a_1 }{c_0} \sim  \frac{c_1 }{c_0}\sim  \frac{c_2 }{c_0}\sim {\mathcal O} (1).
\end{equation}
Therefore the coefficients $a_{0,1}$ and $c_{1,2}$  can be at the same level of
importance as $c_0$ in the modified power counting.
\par
We will take the spin-flip interactions into
account to predict the polarization of $\psi'$ at Tevatron. At Tevatron $\psi'$
can be produced through a color-singlet $^3S_1$ $c\bar c$ pair and
a color-octet $c\bar c$ pair in a $^1S_0$, $^3S_1$ and $^3P_J$ state.
Beside the color-octet $^3S_1$ state,
the spin-flip interactions can also affect the transitions of other states.
Since the color-octet $^3S_1$ state gives the dominant contribution at large
$p_\perp$, we neglect the spin-flip interactions in contributions from other states. Summing all
contributions the cross-section
can be written as:
\begin{equation}
d\sigma = d\sigma [ ^3 S_1^{(1)} ]+ d\sigma [ ^1 S_0^{(8)} ]+d\sigma [ ^3 P_J^{(8)} ]+d\sigma [ ^3 S_1^{(8)} ].
\end{equation}
Among the first three contributions each one
is proportional to a corresponding NRQCD matrix element, which is
$\langle O_1^{\psi'}(^3 S_1) \rangle $, $\langle O_8^{\psi'}(^1 S_0) \rangle $
or $\langle O_8^{\psi'}(^3 P_J) \rangle $. These matrix elements
are determined by fitting the measured cross section of unpolarized
$\psi'$ at certain level. For the contribution $d\sigma [ ^3 S_1^{(8)} ]$,
the matrix $H_{ij}$ can be extracted from the known results in \cite{AKL}.
We also calculated all contributions in the above and find an agreement
with those in \cite{AKL}. Without the spin-flip interactions $d\sigma [ ^3 S_1^{(8)} ]$
is proportional to the matrix element $\langle O_8^{\psi'}(^3 S_1) \rangle $.
Taking the interactions into account,
one can see that the determined $\langle O_8^{\psi'}(^3 S_1) \rangle $ is approximately equal to $3(2c_0
+3 a_0 +a_1)$ from Eq.(4).
\par
\begin{figure}[hbt]
\begin{center}
\includegraphics[width=0.45\textwidth]{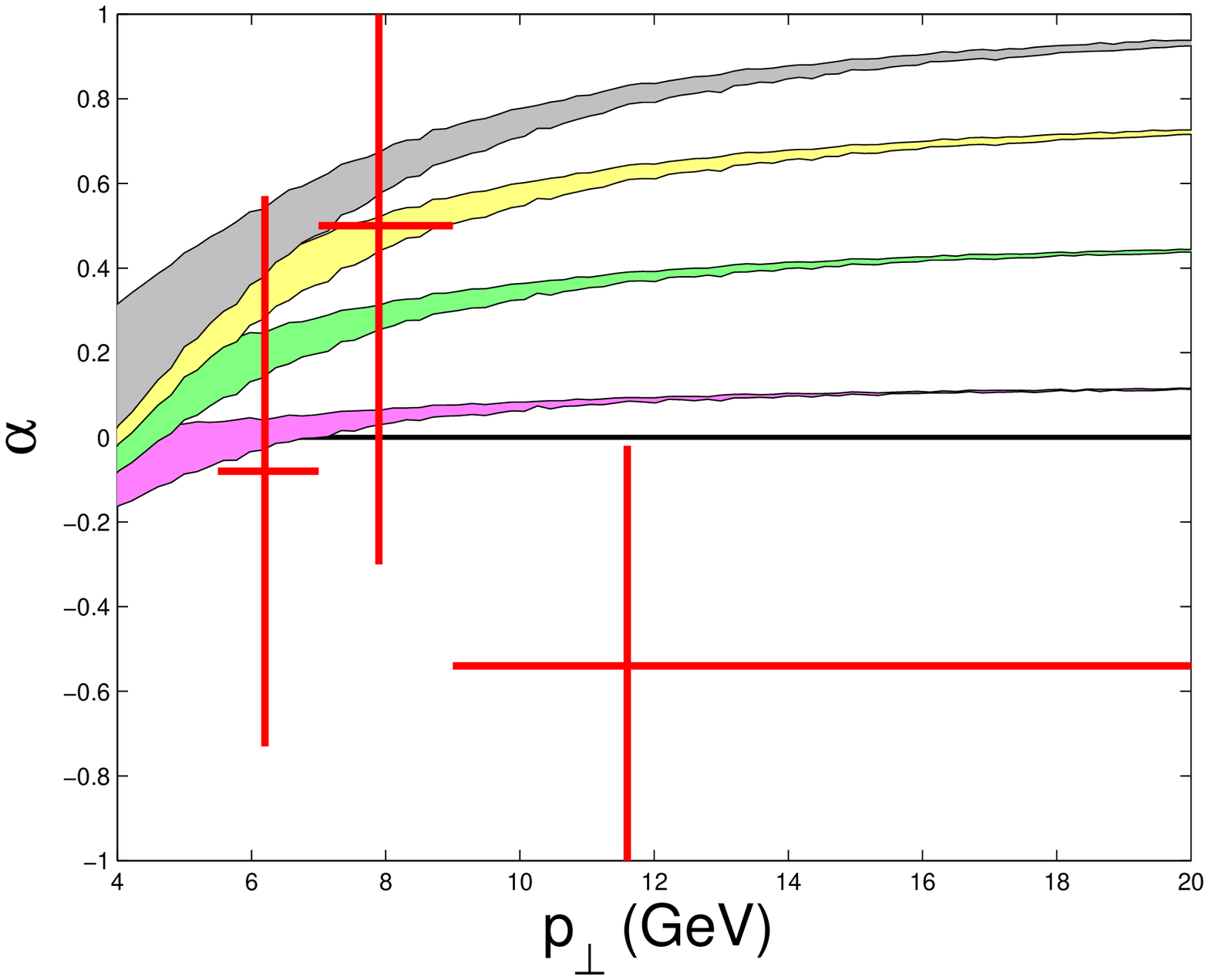}%
\hspace{0.1cm}
\includegraphics[width=0.45\textwidth]{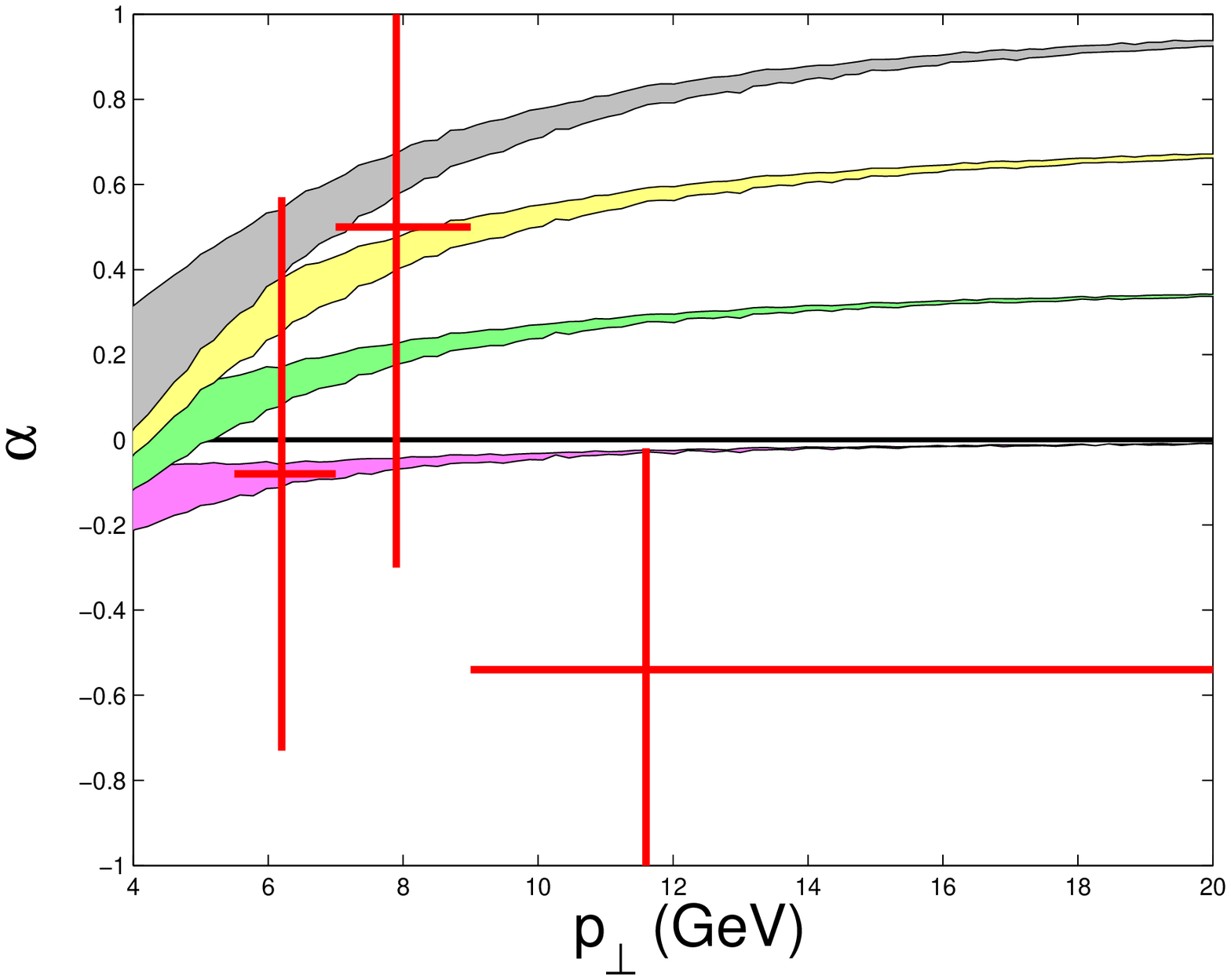}%
\end{center}
\caption{The predicted $\alpha$ as a function
of $p_\perp$. The 4 bands from the top to the bottom are corresponding
to $f_0=0,\ v^4,\ v^2,\ 1$ respectively with $v^2=0.3$.  The area of the bands is obtained
by varying the unknown parameter $x$ from $0$ to $1$. In the left diagram
$f_1$ is taken as $v f_0$. In the right one $f_1$ is taken as $f_0$  }
\label{Feynman-dg1}
\end{figure}
\par
To predict the polarization we make the ansatz that $a_1=c_1=c_2$ and introduce
two parameters $f_0 =a_0/c_0$ and $f_1 = a_1/c_0$. These two parameters
can be from $0$ to order of $1$.
The polarization is predicted
with the parameter $\alpha$ as a function of $p_\perp$, which is defined
as:
\begin{equation}
\alpha =  \left ( \frac{d \sigma_{tot}}{dp_\perp} -3\frac{d \sigma_{L}}{dp_\perp} \right )
              {\Big /} \left ( \frac{d \sigma_{tot}}{dp_\perp} +\frac{d \sigma_{L}}{dp_\perp} \right ).
\end{equation}
If $\alpha=1$, the produced $\psi'$ is transversely polarized. If $\alpha=-1$
the produced $\psi'$ is longitudinally polarized. For numerical predictions
we use the central values of those matrix elements used in \cite{BKL}.
These central values are determined by taking CTEQ5L parton distributions\cite{CTEQ}.
We will also use CTEQ5L parton distributions. With updated parton distributions
these matrix elements need to be re-determined.
We take the charm quark mass $m=1.5$GeV and
the energy scale of $\mu$ as $\mu^2= 4m^2 + p_\perp^2$.
It should be noted that for $\langle O_8^{\psi'}(^1 S_0) \rangle $
and $\langle O_8^{\psi'}(^3 P_0) \rangle $ only the combination
$M_{3.4}=\langle O_8^{\psi'}(^1 S_0) \rangle +3.4\langle O_8^{\psi'}(^3 P_0)/m^2 \rangle $
is determined. We introduce $x=\langle O_8^{\psi'}(^1 S_0) \rangle/M_{3.4}$ and $x$ can be from $0$
to $1$. For different values of $f_0$ and $f_1$ we obtain the numerical results
for $\alpha$ at Tevatron given in Fig.1.
\par
In Fig.1. the experimental data with errors are taken from \cite{Polex}.
From the figure the produced $\psi'$ will be dominantly with transverse polarization
at large $p_\perp$, if one does not take the spin-flip interaction into account, i.e., $f_0=f_1=0$.
Increasing $f_0$ and $f_1$ from $0$, $\alpha$ will be decreased. If one takes
$f_0$ and $f_1$ at the order of $1$, $\alpha$ is close to $0$. This implies
that the $\psi'$ is unpolarized. It should be note that $f_0$ and $f_1$ can be larger
than $1$.
It is clear that the spin-flip interactions can affect the polarization
significantly and $\alpha$ is more closer to the measured from experiment
than that without these interactions.
It should be emphasized that at large $p_\perp$ $\alpha$ is reduced by $50\%$ if
one takes $f_0 = v^2$ as suggested by the standard power counting.
If the power-counting in Eq.(7) is correct,
there will be
a large fraction of produced $\psi'$ with longitudinal polarization
and the fraction can be larger than the fraction with transversal polarization.
Although the large experimental errors prevent from an estimate of $f_0$ and $f_1$,
but one can see that the data favors $f_0\sim 1$. This is unexpected
with the standard power counting, but consistent with the power counting in Eq.(7).
Although we can not determine $f_0$ theoretically, but we can examine other experimental
data beside those from Tevatron to see if $f_0\sim 1$ is allowed.
The polarization of $\psi'$ has been measured
from inclusive $B$-decay into $\psi'$ at CLEO with $\alpha=-0.45\pm 0.30$\cite{CLEO}.
We have made an analysis by adding the effect of spin-flip interactions
to the relevant prediction in \cite{Bpsi}.
We find that the measured $\alpha$ at CLEO does not exclude the possibility
of $f_0\sim 1$. It should be noted that the parameter $\alpha$ of $\psi'$ has been only measured
at Tevatron and CLEO. Based on our work,  we can conclude that the experimental data do not
exclude the possibility of the large spin-flip effect.
\par
In this letter we do not consider the case with $J/\psi$,
because the production of $J/\psi$ is more complicated than that of $\psi'$.
In this case the $J/\psi$ can be produced not only
directly from a parton scattering, but also from decays of higher quarkonium states
produced directly. These states can be $\psi'$ and $\chi_{cJ}$. In order to understand
the impact of spin-flip interactions for polarizations of $J/\psi$, one needs
to understand the same for those higher quarkonium states, like the case with $\psi'$
studied here. But, in general one can expect that the spin-flip interactions
will decrease the polarization of the produced $J/\psi$ at large $p_\perp$.
The study of $J/\psi$'s polarization, including the analysis of transitions of
various $c\bar c$ pairs into $\psi$ with the spin-flip interactions
and the polarization in the inclusive $B$-decay
will be published elsewhere.
\par
To summarize: We have shown that the spin-flip interactions can have a significant impact
on the transition of a $^3S_1$ $c\bar c$ pair into a $J/\psi$ or $\psi'$.
The impact can be parameterized by introducing new parameters in the transition matrix $T$.
If the heavy quark spin symmetry holds, the matrix has only one parameter, denoted as $c_0$.
Because the charm quark is not heavy enough, or $v$ is not small enough,
these new parameters are not small in comparison with $c_0$ and can be at the same size of $c_0$.
These parameters can significantly reduce the polarization parameter $\alpha$ of $\psi'$, measured
at Tevatron. Including the effect of the spin-flip interactions, the
predicted polarization of $\psi'$ will be close to the measured and
the possibility of the large spin-flip effect is not excluded by known experiments.
This provides a way to solve the puzzle discussed at the beginning.
It is also expected that the spin-flip interactions will change the
polarization $J/\psi$ significantly and can provide
a solution to solve the puzzle of $J/\psi$.

\par
\vskip20pt
\noindent
{\bf Acknowledgements}
\par
We would like to thank Prof. X.D. Ji for interesting discussions.
This work is supported by National Nature
Science Foundation of P. R. China(10421503) and China Postdoctoral Science Foundation.

%%%%%%%%%%%%%%%%%%%%%%%%%%%%%%%%%%%%%%%%%%%%%%%%%%%%%%%%%%%%%%%%%%%%%%%%%%%%%%%%%%%%%%%%%%%%%%%%%%%%%%%%%%%


\begin{thebibliography}{99}

\baselineskip10pt

\bibitem{NRQCD} G.T. Bodwin, E. Braaten and G.P. Lapage, Phys. Rev. D51 (1995) 1125, Phys. Rev. D55 (1997)
5853(E).

\bibitem{pv} G.P. Lepage, L. Magnea, C. Nakhleh, U. Magnea and K. Hornbostel, Phys. Rev. D46, (1992), 4052.

\bibitem{Bra} N. Brambilla et al., hep-ph/0412158.

\bibitem{CrosEx} F. Abe et al., CDF Collaboration, Phys. Rev. Lett. {\bf 79}, (1997), 572,
Phys. Rev. Lett. {\bf 79}, (1997), 578,

\bibitem{BrFl} E. Braaten and S. Fleming, Phys. Rev. Lett. {\bf 74}, (1995), 3327.

\bibitem{ChoWise} P. Cho and M. Wise, Phys. Lett. B346, (1995) 129.

\bibitem{BeRo} M. Beneke and I.Z. Rothstein, Phys. Lett. B372, (1996), 157.

\bibitem{Polth1} M. Beneke and I.Z. Rothstein, Phys. Rev. D55, (1997) R5269.

\bibitem{AKL} A. K. Leibovich, Phys. Rev. D56, (1997) 4412.

\bibitem{BKL} E. Braaten, B.A. Kniehl and J. Lee, Phys. Rev. D62, (2000), 094005

\bibitem{Polex} T. Affolder et al., CDF Collaboration, Phys. Rev. Lett. {\bf 85}, (2000) 2886.

\bibitem{GS} G.A. Schuler, Int. J. Mod. Phys. A12, (1997), 3951.

\bibitem{MB} M. Beneke, hep-ph/9703429.

\bibitem{FRL} S. Fleming, I.Z. Rothstein and A.K. Leibovich, Phys. Rev. D64 (2001) 036002.

\bibitem{NQS} G.C. Nayak, J.W. Qiu and G. Sterman, Phys. Lett. B613 (2005) 45-51,
hep-ph/0501235, Phys. Rev. D72 (2005) 114012, hep-ph/0509021.

\bibitem{MaSi} J.P. Ma and Z.G. Si, Phys. Lett. B625 (2005) 67, hep-ph/0506078.

\bibitem{gfj} E. Braaten and T.C. Yuan, Phys. Rev. Lett. {\bf 71} (1993) 1673, J.P. Ma,
Phys. Lett. B332 (1994) 398, Nucl.Phys. B447 (1995) 405,  E. Braaten and J. Lee, Nucl.Phys. B586 (2000) 427,
J. Lee, Phys.Rev. D71 (2005) 094007.

\bibitem{BLS} G.T. Bodwin, J. Lee and  D.K. Sinclair, Phys. Rev. D72 (2005) 014009.

\bibitem{MNS} G. Martinelli, M. Neubert and C.T. Sachrajda, Nucl. Phys. B461 (1996) 238.

\bibitem{ReCo}
M.~Gremm and A.~Kapustin,
Phys.\ Lett.\ B {\bf 407} (1997) 323,
G.~T.~Bodwin and J.~Lee,
Phys.\ Rev.\ D69 ,(2004) 054003,
G.~T.~Bodwin and A.~Petrelli,
Phys.\ Rev.\ D66 (2002) 094011,
J.~P.~Ma and Q.~Wang,
Phys.\ Lett.\ B {\bf 537} (2002) 233,
C.~B.~Paranavitane, B.~H.~J.~McKellar and J.~P.~Ma,
Phys.\ Rev.\ D61 (2000) 114502,
J.~P.~Ma,
Phys.\ Rev.\ D62  (2000) 054012.

\bibitem{CTEQ} H.L. Lai {\it et. al.},   CTEQ Collaboration,  Eur. Phys. J. C12,
375(2000).

\bibitem{CLEO} S. Anderson {\it et. al.}, CLEO Collaboration, Phys. Rev. Lett. {\bf 89} (2002) 282001.

\bibitem{Bpsi} S. Fleming {\it et. al.}, Phys. Rev. D55 (1997) 4098, P. Ko, J. Lee and H.S. Song, J. Korean
Phys. Soc. {\bf 34} (1999) 301.








\end{thebibliography}
\end{document}